%% file: main.tex
\documentclass[10pt, conference, letterpaper]{IEEEtran}
\IEEEoverridecommandlockouts
\usepackage{amsmath,amsfonts}
\usepackage{amssymb}
\usepackage{algorithmic}
\usepackage{algorithm}
\usepackage{array}
\usepackage{textcomp}
\usepackage{stfloats}
\usepackage{color}
\usepackage{url}
\usepackage{verbatim}
\usepackage{graphicx}
\usepackage{epstopdf}
\usepackage{etoolbox}
\usepackage{cite}
\usepackage{bm}
\usepackage[absolute,overlay]{textpos}
\usepackage{ragged2e}
\begin{document}
\title{Deep Learning-Based Extended Target \\ Tracking in ISAC Systems}

\begin{textblock}{160}(25,5)
\begin{minipage}[t]{\textwidth}
\fontsize{10.7}{12}\selectfont 
\justifying
\noindent
Y. Wang, M. Tao, and S. Sun, “Deep Learning-Based Extended Target Tracking in ISAC Systems,” \textit{2025 IEEE International Conference on Communications Workshops (ICC Wkshps)}, Montreal, Canada, 2025.
\end{minipage}
\end{textblock}

\makeatletter
\newcommand{\linebreakand}{%
  \end{@IEEEauthorhalign}
  \hfill\mbox{}\par
  \mbox{}\hfill\begin{@IEEEauthorhalign}
}
\makeatother
\author{Yiqiu Wang, Meixia Tao, and Shu Sun\\
Department of Electronic Engineering, Shanghai Jiao Tong University, Shanghai, China \\
Emails: \{wyq18962080590, mxtao, shusun\}@sjtu.edu.cn
\vspace{-1em}
\thanks{\textcolor{black}{This work is supported by the Natural Science Foundation of China under Grant 62125108, Grant 62431014, and Grant 62271310, and in part by the Science and Technology Commission Foundation of Shanghai under Grant 24DP1500702.}}}

\maketitle
\begin{abstract}
In this paper, we explore the feasibility of using communication signals for extended target (ET) tracking in an integrated sensing and communication (ISAC) system. The ET is characterized by its center range, azimuth, orientation, and contour shape, for which conventional scatterer-based tracking algorithms are hardly feasible due to the limited scatterer resolution in ISAC. To address this challenge, we propose ISACTrackNet, a deep learning-based tracking model that directly estimates ET kinematic and contour parameters from noisy received echoes. The model consists of three modules: Denoising module for clutter and self-interference suppression, Encoder module for instantaneous state estimation, and KalmanNet module for prediction refinement within a constant-velocity state-space model. Simulation results show that ISACTrackNet achieves near-optimal accuracy in position and angle estimation compared to radar-based tracking methods, even under limited measurement resolution and partial occlusions, but orientation and contour shape estimation remains slightly suboptimal. These results clearly demonstrate the feasibility of using communication-only signals for reliable ET tracking.
\end{abstract}

\section{Introduction}
\IEEEPARstart{T}{arget} tracking refers to determining the kinematic states of specific objects within a surveillance area based on sensor measurements. In its most typical formulation, each object is modeled as a \textit{point target}, and the sensor can generate at most one measurement near the object's center per scan. This leads to the most simplified target tracking task, where we only need to estimate the position (and optionally the velocity) of a given object. With the rapid development of high-resolution radar, each target can now generate several measurements along its contour, where multiple scatterers on the same object can be resolved by radar sensors. This has led to an important and challenging problem of exploring scatterer-to-target geometry, namely extended target (ET) tracking.

Comprehensive characterization of an ET requires real-time estimation of the target's position, orientation, along with its contour shape. The contour modeling of an ET typically employs either basic geometric shapes (e.g., rectangles and ellipses), or more sophisticated irregular shapes. In the radar literature, extensive ET tracking algorithms have been developed based on these contour models. Notably, the work \cite{Koch08TAES} introduces a random matrix model for elliptical targets, where the unknown elliptical contour is represented by a symmetric and positive definite matrix and then tracked and updated via a Kalman-filter-like method. Nevertheless, the original random matrix work \cite{Koch08TAES} uses an implicit assumption that the measurement noise must be proportional to the target contour, which restricts its applications. This problem is later addressed in \cite{Feldmann11TSP} by utilizing uniform distribution for measurement likelihood, which also proved that such distribution can be well approximated by a refined Gaussian distribution. The above results are experimentally verified in \cite{Vivone17JAIF}. For targets with irregular contour, multiple sub-ellipses model is proposed in \cite{Lan12FUSION}. Aside from ellipse-based contour, another tracking scheme models the target contour with parametric star-convex shapes. The work in \cite{Wahlström15TSP} uses Gaussian Process (GP) for modeling the target contour. Each part of the unknown contour is learned and updated via an extended Kalman Filter. In highly dynamic environments, a refined particle filter-based scheme is proposed as a robust tracker for the same model \cite{Emre16FUSION}.

While the aforementioned radar-based techniques for ET tracking are well investigated, ET tracking in integrated sensing and communication (ISAC) systems remains largely unexplored. As a key enabling technology for 6G networks, ISAC is expected to enable dual functions of communication and sensing based on a shared use of spectrum, waveform, and hardware. In other words, ISAC allows the existing communication signals to be specifically repurposed for sensing tasks, without sending dedicated radar signals. Our previous works \cite{Wang24TWC,Wang24arxiv} investigate the monostatic sensing for static ETs with arbitrary shapes in ISAC systems. The work \cite{Du23TWC} considers the tracking of a vehicle target where multiple resolvable scatterers along the ET are tracked from base station (BS) communication echoes. This is later extended to the aerial scenario in \cite{Pang24TWC} where a flying drone provides tracking service for ground ETs. Note that the above works \cite{Du23TWC,Pang24TWC} follow an echo-to-scatterer paradigm where the ET tracking task essentially degrades to the tracking of separated scatterers requiring an extra scatterer-to-target step. However, unlike radar systems, the limited spectrum and power resources in communication signals result in poor angle and range resolution for echo-to-scatterer mapping. The noisy and sparsely resolved scatterers from communication echoes would further restrict the performance of scatterer-to-target filters.

To address the above challenges, this work aims to develop a super-resolution ISAC-based ET tracking scheme that bypasses the traditional echo-to-scatterer step and establishes a direct echo-to-target paradigm, which infers the ET kinematic and shape parameters from noisy communication echoes. 
To this end, we propose a \textit{three-stage deep-learning} based tracking model, \textbf{ISACTrackNet}, that integrates an advanced Denoising module, an instantaneous state Encoder module, and an iterative KalmanNet filtering module for accurate ET state estimation. 
Specifically, the \textit{Denoising module} suppresses clutter echo and residual self-interference (SI) embedded in received signals via an autoencoder-style network, ensuring high-fidelity ET echo extraction.
On top of that, the \textit{Encoder module} converts the de-noised echo into instantaneous kinematic and shape estimations by leveraging historical ET channel features.
Finally, the \textit{KalmanNet} module learns system uncertainties through a GRU-based mechanism, refining the coarse estimations in an iterative manner within a state-space model.
By synergizing these three modules, ISACTrackNet circumvents the scatterer-resolution barrier in ISAC systems, and demonstrating near-optimal accuracy in tracking the ET center range, azimuth, orientation, and shape parameters.

\textit{Notations:} $\left[\cdot\right]^T$, $\left[\cdot\right]^H$ denote, respectively, the transpose and Hermitian transpose of a matrix; $\mathbb{E}\left[\cdot\right]$ denotes the averaging operation; $\mathcal{CN}\left(\mathbf{0}_{m\times 1},\sigma^{2}\mathbf{I}_m\right)$ denotes the probability density function of an ${m\times 1}$ circularly symmetric complex Gaussian vector with zero mean and covariance $\sigma^{2}\mathbf{I}_m$; $\Re\left(\cdot\right)$ and $\Im\left(\cdot\right)$ denote the real and imaginary part of a complex number; $\mathbb{R}^{m\times n}$ and $\mathbb{C}^{m\times n}$ denote a matrix with ${m\times n}$ real and complex elements; \textcolor{black}{$\rm{diag}\left[\mathbf{A}, \mathbf{B}\right]$ denotes a block diagonal matrix with $\mathbf{A}$ and $\mathbf{B}$ as diagonal blocks}; $\lVert \cdot \rVert$ denotes the $L_2$ norm of a vector.

\begin{figure}[!t]
\centering
\includegraphics[width=3in]{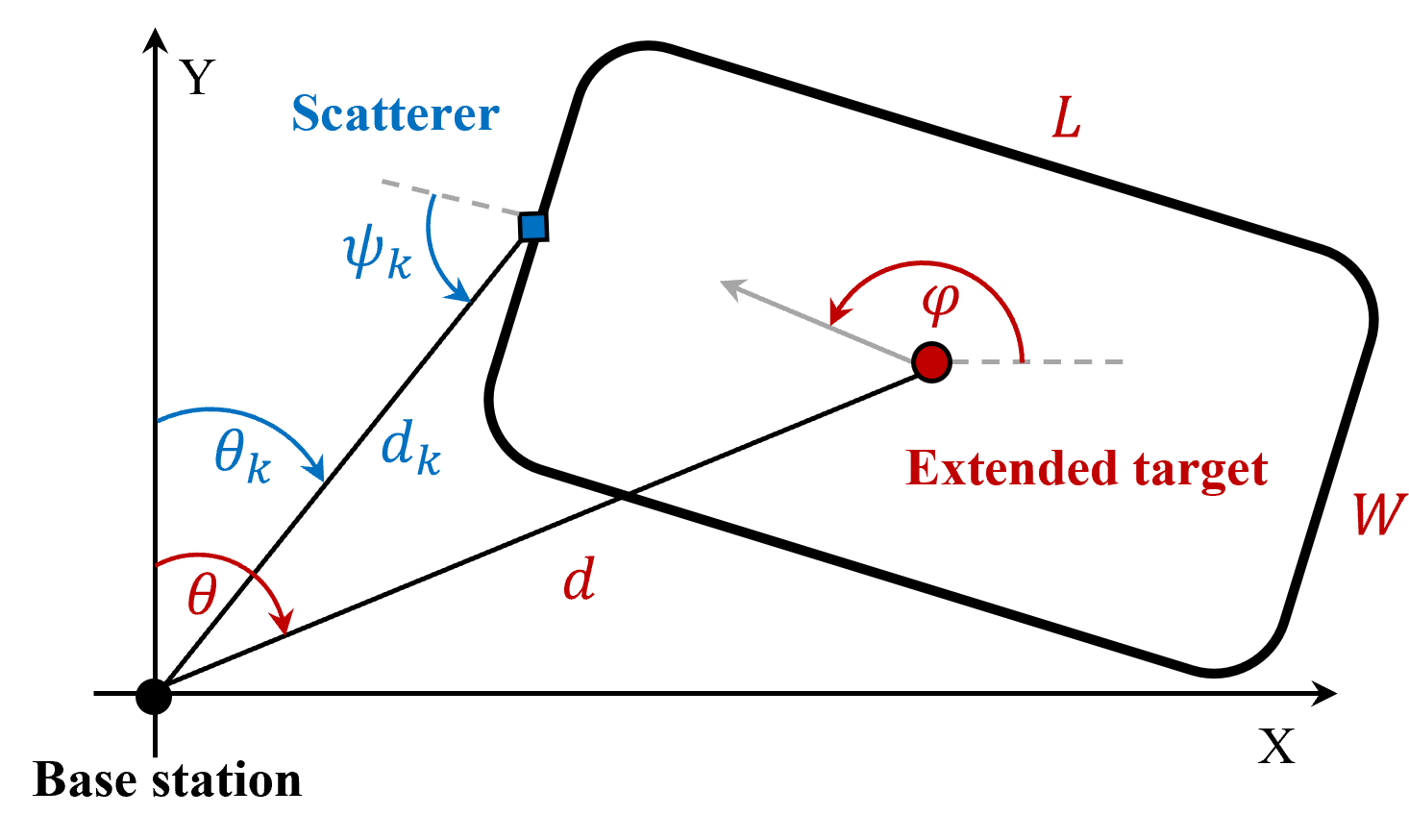}
\caption{The rectangular extended target in the global coordinate. The base station is located at the origin. The red and blue markers refer to the ET center and a visible scatterer along the ET contour, respectively.}
\label{fig_ET}
\end{figure}

\section{System Model}
In this paper, we consider a downlink ISAC system where a full-duplex BS sends communication symbols and collects back-scattered echoes for ET tracking. The BS is equipped with one pair of uniform linear arrays (ULAs) with $N_t$ transmit antennas and $N_r$ receive antennas respectively. It is assumed that the ET has a rectangular contour but with unknown size, and multiple scatterers are randomly distributed along the contour.
\subsection{Transmit Signal Model}
We investigate a fully integrated ISAC scenario in which the BS exclusively utilizes communication symbols for sensing, without transmitting any dedicated radar probing signals. Let $\mathbf{x}_n \in \mathbb{C}^ {N_t}$ denote the transmit signal from BS at time $n$ as
\begin{equation}
\label{x_n}
\mathbf{x}_n = \mathbf{w}_n s_n,
\end{equation}
where $\mathbf{w}_n \in \mathbb{C}^{N_{t}}$ is the transmit beamforming vector, and $s_n$ is the information symbol. The information symbols are normalized with unit power as $\mathbb{E} \left[s_n s^*_n \right] = 1$.

To facilitate ET tracking, the transmit beamformer $\mathbf{w}_n$ should be designed to illuminate the whole contour of the target based on past tracking results. In other words, the half-power width of the steered beam is required to be sufficient to cover the entire target in physical size. Following \cite{Du23TWC}, here we adopt a practical analog beamforming approach, which utilizes the steering vector towards the predicted ET center azimuth $\bar{\theta}_{n|n-1}$ the transmit beamformer, given by
\begin{equation}
\mathbf{w}_n = [\mathbf{a}(\bar{\theta}_{n|n-1},N_{t,n})^T,\mathbf{0}_{1\times{(N_t-N_{t,n})}}]^T.\label{beamformer}
\end{equation}
Here, $\mathbf{a}(\cdot)$ is the transmit steering vector, and $N_{t,n}$ is the number of activated transmit antennas, adjustable for dynamic beamforming with varying beamwidth at different time slots. The proper value for $N_{t,n}$ will be further discussed in Section III-C. The transmit steering vector $\mathbf{a}(\theta,N_{t,n})$ and receive steering vector $\mathbf{b}(\theta,N_r)$ are respectively defined as 
\begin{align}
    &\mathbf{a}(\theta,N_{t,n}) = \frac{1}{\sqrt{N_{t,n}}}\left[1,e^{-j\pi\sin\theta},...,e^{-j\pi(N_{t,n}-1)\sin\theta}\right]^T,\\
    &\mathbf{b}(\theta,N_r) = \frac{1}{\sqrt{N_r}}\left[1,e^{-j\pi\sin\theta},...,e^{-j\pi(N_r-1)\sin\theta}\right]^T.
\end{align}
\subsection{Received Sensing Signal Model}
At the $n$-th time slot, the BS receives the sensing signal $\mathbf{y}_n\in \mathbb{C}^{N_r}$ with multiple components \cite{He23JSAC}, expressed as
\begin{equation}
\label{sensing signal}
    \mathbf{y}_n=\mathbf{e}_n+\mathbf{c}_n+\mathbf{r}_n+\mathbf{z}_n,
\end{equation}
where $\mathbf{e}_n\in \mathbb{C}^{{N}_{r}}$ and $\mathbf{c}_n\in \mathbb{C}^{{N}_{r}}$ are respectively the echo signals scattered from the ET and environment clutters, $\mathbf{r}_n\in \mathbb{C}^{{N}_{r}}$ is the residual transmitter SI caused by full-duplex operation, and $\mathbf{z}_n\sim \mathcal{CN}(0, \sigma^2 \mathbf{I}_{N_r})$ is the additive sensing noise. Next, we will discuss the detailed modeling of the above sensing signal components.

\subsection{Extended Target Echo}
As shown in Fig.~\ref{fig_ET}, in this work we aim to track a rectangular ET located in the $Oxy$ plane. The ET echoes are considered as signals scattered from the visible elements along the ET contour, whereas the energy generated from internal reflection is negligible due to the severe penetration loss. Define $\mathcal{C}$ as the visible part of the ET contour (from the BS viewpoint). We can divide the visible contour into $K$ nonoverlapping regions satisfying ${\mathcal{C}}=\bigcup_{k=1}^{K}{\mathcal{C}_{k}}$ and ${\mathcal{C}_{k_1}}\bigcap{\mathcal{C}_{k_2}}=\varnothing,\forall {k_1}\neq{k_2}$. For simplicity, we use \textit{scatterer} to represent each divided contour section. The received echo signal at time $n$ \cite{Wang24TWC} is modeled as
\begin{align}
&\mathbf{e}_n=\int_{\mathcal{C}}{\mathbf{e}}_{n,\boldsymbol{\rho}}\text{d}\boldsymbol{\rho}\approx\sum\nolimits_{k=1}^{K}{{\mathbf{e}}_{n,k}},\\
&{\mathbf{e}}_{n,k}=g_{n,k}\sqrt{l_{n,k}}\zeta_{n,k}\mathbf{b}_{n,k}\mathbf{a}^{H}_{n,k}\mathbf{x}_n,\label{e_k}
\end{align}
where ${{\mathbf{e}}_{n,\boldsymbol{\rho}}}$ and $\mathbf{e}_{n,k}$ refer to the echo signals as a function of visible elements and $\mathcal{C}_k$, ${\zeta }_{n,k}=(\cos\psi_{n,k})^2$, $\psi_{n,k}$, ${\theta}_{n,k}$, $d_{n,k}$, and $l_{n,k}$ refer to the radar cross section (RCS), angle between the contour normal and scatterer-BS path, global azimuth angle, range, and equivalent length of the $k$-th scatterer, respectively. Here $k\in\mathcal{K}$ and $\mathcal{K}=\{1,...,K\}$ is the set of ET scatterers, $g_{n,k}=\sqrt{p_0}/d_{n,k}^{2}$ is the sensing path loss coefficient, $p_0$ is the reference path loss at $1\ \mathrm{m}$ distance, and $\mathbf{a}_{n,k}$, $\mathbf{b}_{n,k}$ are abbreviations for $\mathbf{a}\left(\theta_{n,k},N_t\right)$ and $\mathbf{b}\left(\theta_{n,k},N_r\right)$.

\subsection{Clutter Echo}
The environment clutters are assumed to be static scatterers randomly distributed in the surveillance area of the BS. The undesired clutter echo signal at time $n$ is modeled as \cite{He23JSAC}
\begin{equation}
    \mathbf{c}_{n}=\sum\nolimits_{k=1}^{K_\mathrm{CL}}g_{n,k}\bar{\zeta}_{n,k}\mathbf{b}_{n,k}\mathbf{a}^{H}_{n,k}\mathbf{x}_n,
\end{equation}
where $\bar{\zeta}_{n,k} \sim \mathcal{CN}(0,1)$ is the RCS of the $k$-th clutter, $K_{\rm{CL}}$ is the number of clutters, and $g_{n,k}$, $\mathbf{b}_{n,k}$, $\mathbf{a}_{n,k}$ share the same definition as in \eqref{e_k}.

\subsection{Residual Self-Interference}
With the application of conventional analog and digital SI cancellation techniques, the residual SI $\mathbf{r}_n$ at time $n$ can be suppressed to a relatively low energy level. Here we consider the suppressed near-field signal leakage from the transmitter as the residual SI \cite{Satyanarayana19TVT}, written as
\begin{equation}
\mathbf{r}_n=\mathbf{H}_{\mathrm{SI}}\mathbf{x}_n,\ \ \ \mathbf{H}_{\mathrm{SI}}^{(i,j)} = \frac{p_{\mathrm{SI}}}{r^{(i,j)}}\exp{\left(-j2\pi{r^{(i,j)}}/{\lambda}\right)},
\end{equation}
where $\mathbf{H}_\mathrm{SI}$ and $p_\mathrm{SI}$ are the channel and average power of residual SI, $\mathbf{H}_\mathrm{SI}^{(i,j)}$ refers to the element at the $i$-th row and $j$-th column of $\mathbf{H}_\mathrm{SI}$, $r^{(i,j)}$ is the distance between the $i$-th transmit antenna and the $j$-th receive antenna, and $\lambda$ refers to the signal wavelength.

\section{Problem Formulation of ET Tracking}
We aim to directly track the ET kinematic and shape parameters at each time $n$ from the historically transmitted signal sequence ${\mathbf{x}_{1:n}}$ and the historically received signal sequence ${\mathbf{y}_{1:n}}$, following the echo-to-target paradigm. Here the subscript $_{1:n}$ refers to the combination of sequences from slot $1$ to slot $n$. Let $\boldsymbol{\Theta}_n=[x_n,y_n,\phi_n,L,W]^T$ denote ET kinematic parameters to be tracked, which include the ET center position $[x_n,y_n]$, orientation $\phi_n$, and the fixed shape attributes (length $L$ and width $W$). The general tracking problem can be formulated as
\begin{align}
 \hat{\boldsymbol{\Theta}}_n= f(\mathbf{x}_{1:n},\mathbf{y}_{1:n}),\label{problem formulation}
\end{align}
\subsection{Deep Learning-based ET Tracking}
As discussed earlier, the ET tracking task in ISAC faces an implicit echo-to-target mapping issue, which can hardly be solved by traditional model-based tracking algorithms. Thus, here we adopt a deep learning technique and propose ISACTrackNet to capture the implicit echo-to-target mapping in \eqref{problem formulation}. As illustrated in Fig.~\ref{fig_NetworkStructure}, the proposed ISACTrackNet includes three modules, described as follows:
\begin{figure*}[!t]
\centering
\includegraphics[width=6in]{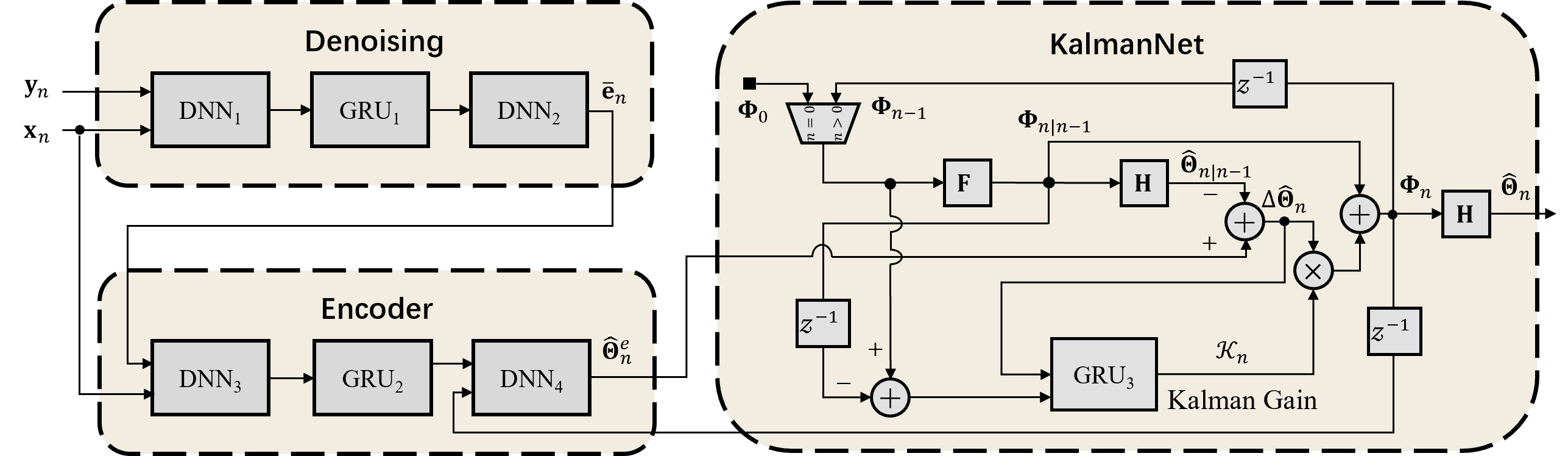}
\caption{The structure of the proposed ISACTrackNet. \textcolor{black}{The adopted DNNs have different sizes, with $\rm{DNN}_1$: $[2N_t+2N_r,\ 8N_t+8N_r,\ 4N_t+4N_r,\ 2N_t+2N_r]$, $\rm{DNN}_2$: $[2N_t+2N_r,\ 8N_r,\ 4N_r,\ 2N_r]$, $\rm{DNN}_3$: $[2N_t+2N_r,\ 128,\ 64,\ 32]$, and $\rm{DNN}_4$: $[40,\ 64,\ 32,\ 16]$. In $\rm{DNN}_4$, external inputs from KalmanNet are introduced in the first layer. The hidden dimensions of $\rm{GRU}_1$ and $\rm{GRU}_2$ are respectively $2N_t+2N_r$ and 32. $\rm{GRU}_3$ uses architecture \#2 in \cite{Revach22TSP}.}}
\label{fig_NetworkStructure}
\end{figure*}

1) \textit{Denoising Module}: 
This module follows an autoencoder-style framework, employing two DNNs and one GRU to extract the de-noised ET echo signal. The first DNN performs dimensionality reduction on the concatenated received signal $\mathbf{y}_n$ and transmitted signal $\mathbf{x}_n$, yielding a low-dimensional feature representation. The GRU then leverages historical feature vectors (from time slot $1$ to $n-1$) to progressively strip away clutter echo and residual self-interference components from the received signal. Finally, the second DNN reconstructs the de-noised echo $\bar{\mathbf{e}}_n$ from GRU-refined features. The whole process is represented as
\begin{equation}
    \bar{\mathbf{e}}_n = f_1(\mathbf{x}_n,\mathbf{y}_n).
\end{equation}

2) \textit{Encoder Module}: This module also utilizes two DNNs and one GRU to estimate the ET kinematic and shape parameters at each time slot. The first DNN extracts channel features from both the de-noised echo $\bar{\mathbf{e}}_n$ and the transmitted signal $\mathbf{x}_n$. Next, the GRU incorporates past channel information (from time slot $1$ to $n-1$) to refine these features for the current time step. In the second DNN, external predictive parameters $\boldsymbol{\Phi}_{n-1}$ from KalmanNet are introduced as prior knowledge and jointly processed with the refined channel features, enabling a coarse estimation of the current target state $\hat{\boldsymbol{\Theta}}_{n}^{e} = [\hat{x}_{n}^e,\hat{y}_{n}^e,\hat{\phi}_n^e,\hat{L}_n^e,\hat{W}_n^e]$. The encoder process is written as
\begin{equation}
    \hat{\boldsymbol{\Theta}}_n^e = f_2(\mathbf{x}_n,\bar{\mathbf{e}}_n,\boldsymbol{\Phi}_{n-1}).
\end{equation}

3) \textit{KalmanNet Module}:  
While the rough tracking result $\hat{\boldsymbol{\Theta}}_n^e$ can be further refined with a Kalman Filter, it is generally difficult to obtain the covariance matrices of the process and measurement noise in conventional Kalman Filters, as they are often manually tuned or empirically determined. To address this, a learning-based KalmanNet was introduced in \cite{Revach22TSP}, where the unknown covariance matrices are learned and updated based on historical tracking and measurement results.

Once the parameter estimate $\hat{\boldsymbol{\Theta}}_n^e$ is obtained from the Encoder module, it is treated as the external measurement input to KalmanNet at time $n$. The state prediction is then computed based on the state evolution model, while the Kalman gain is adaptively generated via a GRU module. These components are integrated into a standard Kalman filtering pipeline to yield the final refined estimate of the target state $\hat{\boldsymbol{\Theta}}_n$.

Here, we adopt a \textit{constant velocity}\footnote{In most tracking applications, the target always experiences non-constant-velocity movement. Thus, the proposed ISACTrackNet is required to be robust enough to handle minor state-space model mismatches.} state-space model \cite{Wahlström15TSP} in KalmanNet, defined as:
\begin{align}
    \text{\small State Evolution Model: } &\boldsymbol{\Phi}_n = \mathbf{F}\boldsymbol{\Phi}_{n-1} + \boldsymbol{\nu}_n,\label{state-evolution}\\
    \text{\small Measurement Model: } &\hat{\boldsymbol{\Theta}}_n^e = \mathbf{H}\boldsymbol{\Phi}_{n} + \hat{\boldsymbol{\nu}}_n,\\
    \text{\small Evolution Matrix: } &\mathbf{F} = \mathrm{diag}\Bigl({\begin{bmatrix}
        1 & T \\ 0 & 1 \end{bmatrix} \otimes \mathbf{I}_3, \mathbf{I}_2}\Bigr),\\
    \text{\small Measurement Matrix: } &\mathbf{H} = \mathrm{diag}\Bigl(\mathbf{I}_3,\bigl[\mathbf{0}_{2\times3},\mathbf{I}_2\bigr]\Bigr),
\end{align}
where $\boldsymbol{\nu}_n$ and $\hat{\boldsymbol{\nu}}_n$ represent the process and observation noise vectors arising from model uncertainties and sensing imperfections. The tracked state vector in KalmanNet is defined as $\boldsymbol{\Phi}_n = [\bar{x}_{n},\bar{y}_{n},\bar{\phi}_n,\bar{v}_{x,n},\bar{v}_{y,n},\bar{\varphi}_{n},\bar{L}_n,\bar{W}_n]^T$, and $T$ is the observation interval. The velocity terms ($\bar{v}_{x,n}, \bar{v}_{y,n}, \bar{\varphi}_{n}$) are additionally introduced to support the constant velocity model.

The KalmanNet filtering process is expressed as:
\begin{equation}
     \hat{\boldsymbol{\Theta}}_n = \mathbf{H}\boldsymbol{\Phi}_n,\ \ \ \boldsymbol{\Phi}_n = f_3(\hat{\boldsymbol{\Theta}}_n^e,\boldsymbol{\Phi}_{n-1},\boldsymbol{\Phi}_0,\mathbf{F},\mathbf{H}),
\end{equation}
where $\boldsymbol{\Phi}_0$ denotes the initial ET state.

\subsection{Training Strategy and Loss Function Design}
Considering the functionalities of the different modules in ISACTrackNet, we separate the whole training process into four stages, each with its unique loss functions.

1) \textit{Denoising Module Training}: In this phase, only the Denoising module is trained to eliminate undesired residual SI and clutter echoes from the received signal, while all other modules remain untrained. The loss function is designed as
\begin{equation}
   \mathcal{L}_1 = \frac{1}{MN}\sum\nolimits_{m=1}^{M}\sum\nolimits_{n=1}^N(1-e^{-\alpha n})\lVert\mathbf{e}_n-\bar{\mathbf{e}}_n\rVert^2,\label{L2}
\end{equation}
where $M$ and $N$ are respectively the numbers of training samples and time slots, and $\alpha$ is a forgetting factor which places higher weight on the tracking performance of later timestamps.

2) \textit{Encoder Module Training}: In this phase, we aim to achieve a warm start for instantaneous target estimation without KalmanNet. Thus, only the Encoder module is trained in this phase. The loss function is written as
\begin{equation}
       \mathcal{L}_2 = \frac{1}{MN}\sum\nolimits_{m=1}^{M}\sum\nolimits_{n=1}^N(1-e^{-\alpha n})\lVert\boldsymbol{\Theta}_n-\hat{\boldsymbol{\Theta}}_n^e\rVert^2.\label{L2}
\end{equation}

3) \textit{Encoder + KalmanNet Module Training}: With a well pre-trained Encoder module, we now combine KalmanNet together with Encoder for better tracking performance. Here, the parameters of KalmanNet and Encoder modules are updated in an alternating manner. Specifically, KalmanNet is first updated while keeping Encoder fixed, and then Encoder is updated while keeping KalmanNet fixed. This process is repeated until convergence. The loss function is expressed as
\begin{equation}
       \mathcal{L}_3 = \frac{1}{MN}\sum\nolimits_{m=1}^{M}\sum\nolimits_{n=1}^N(1-e^{-\alpha n})\lVert\boldsymbol{\Theta}_n-\hat{\boldsymbol{\Theta}}_n\rVert^2.\label{L3}
\end{equation}

4) \textit{Joint Training}: All modules get trained in this phase for final update. The loss function is same with \eqref{L3}.

\subsection{Beamforming Design}
To facilitate ET tracking, recall that we have designed an analog beamformer $\mathbf{a}(\bar{\theta}_{n|n-1},N_{t,n})$ back in \eqref{beamformer} with sufficient beamwidth to cover the whole ET. According to the discussion in Section III-A, the predicted ET center azimuth $\bar{\theta}_{n|n-1}$ in KalmanNet is now written as
\begin{equation}
\bar{\theta}_{n|n-1} =\mathrm{atan}{\left(\frac{\bar{x}_{n|n-1}}{\bar{y}_{n|n-1}}\right)}= \mathrm{atan}{\left(\frac{\bar{x}_{n-1}+T\bar{v}_{x,n-1}}{\bar{y}_{n-1}+T\bar{v}_{y,n-1}}\right)}.\label{predicted azimuth}
\end{equation}

Next we need to determine the proper value for $N_{t,n}$, the number of activated transmit antennas. The commonly used beamwidth for a ULA is approximated by \cite{Du23TWC}
\begin{align}
\theta_\mathrm{BW} \approx {1.78}({N_{t,n} \cos\bar{\theta}_{n|n-1}})^{-1}.
\end{align}

The coverage width $\Delta d$ can thus be approximately calculated using trigonometric functions as
\begin{align}
\Delta d \approx 2\bar{d}_{n|n-1} \cdot \tan\left(\frac{0.89}{N_{t,n} \cos\bar{\theta}_{n|n-1}}\right),
\end{align}
where $\bar{d}_{n|n-1} =\sqrt{\bar{x}_{n|n-1}^2+\bar{y}_{n|n-1}^2}$ is the predicted distance.

In our considered scenario, the beam coverage width should always be larger than the diagonal length of the ET, namely $\Delta d^2 > \bar{L}_{n-1}^2+\bar{W}_{n-1}^2$. Thus, the number of activated antenna elements in the analog beamformer should be adjusted accordingly for full coverage of the ET:
\begin{align}
    N_{t,n} = \min \Biggl\{ \Bigg\lfloor\frac{0.89}{\mathrm{atan} \Bigl( \frac{\sqrt{\bar{L}_{n-1}^2+\bar{W}_{n-1}^2}}{2 \bar{d}_{n|n-1}} \Bigr) \cos \Bigl( \bar{\theta}_{n|n-1} \Bigr)}\Bigg\rfloor, N_{t} \Biggr\}.\label{activated antenna}
\end{align}

The overall tracking algorithm is outlined in Algorithm 1.
\begin{algorithm}[!h]
    \caption{Deep Learning-based Tracking Algorithm}
    \label{alg:Tracking}
    \renewcommand{\algorithmicrequire}{\textbf{Input:}}
    \renewcommand{\algorithmicensure}{\textbf{Output:}}
    \begin{algorithmic}[1]
        \REQUIRE $\mathbf{s}_{1:N}$, $\boldsymbol{\Phi}_0$, $\mathbf{F}$, $\mathbf{H}$  \ENSURE $\hat{\boldsymbol{\Theta}}_{1:N}$
        
        \STATE  Initialize: $n=1$.
        \WHILE{$n < N$}
            \STATE $n = n + 1.$
            \STATE Determine the predicted ET azimuth $\bar{\theta}_{n|n-1}$ and activated antenna number $N_{t,n}$ in \eqref{predicted azimuth} and \eqref{activated antenna}.
            \STATE Generate beamformer $\mathbf{w}_n$ in \eqref{beamformer}.
            \STATE Generate transmitted signal $\mathbf{x}_n$ in \eqref{x_n}.
            \STATE Receive noised echo signal $\mathbf{y}_n$ in \eqref{sensing signal}.
            \STATE Obtain ISACTrackNet result $\hat{\boldsymbol{\Theta}}_n$ in \eqref{problem formulation}.
        \ENDWHILE
        \STATE Cascade all tracking results and obtain $\hat{\boldsymbol{\Theta}}_{1:N}$.
    \end{algorithmic}
\end{algorithm}
\begin{figure}[!t]
\centering
\includegraphics[width=3.58in]{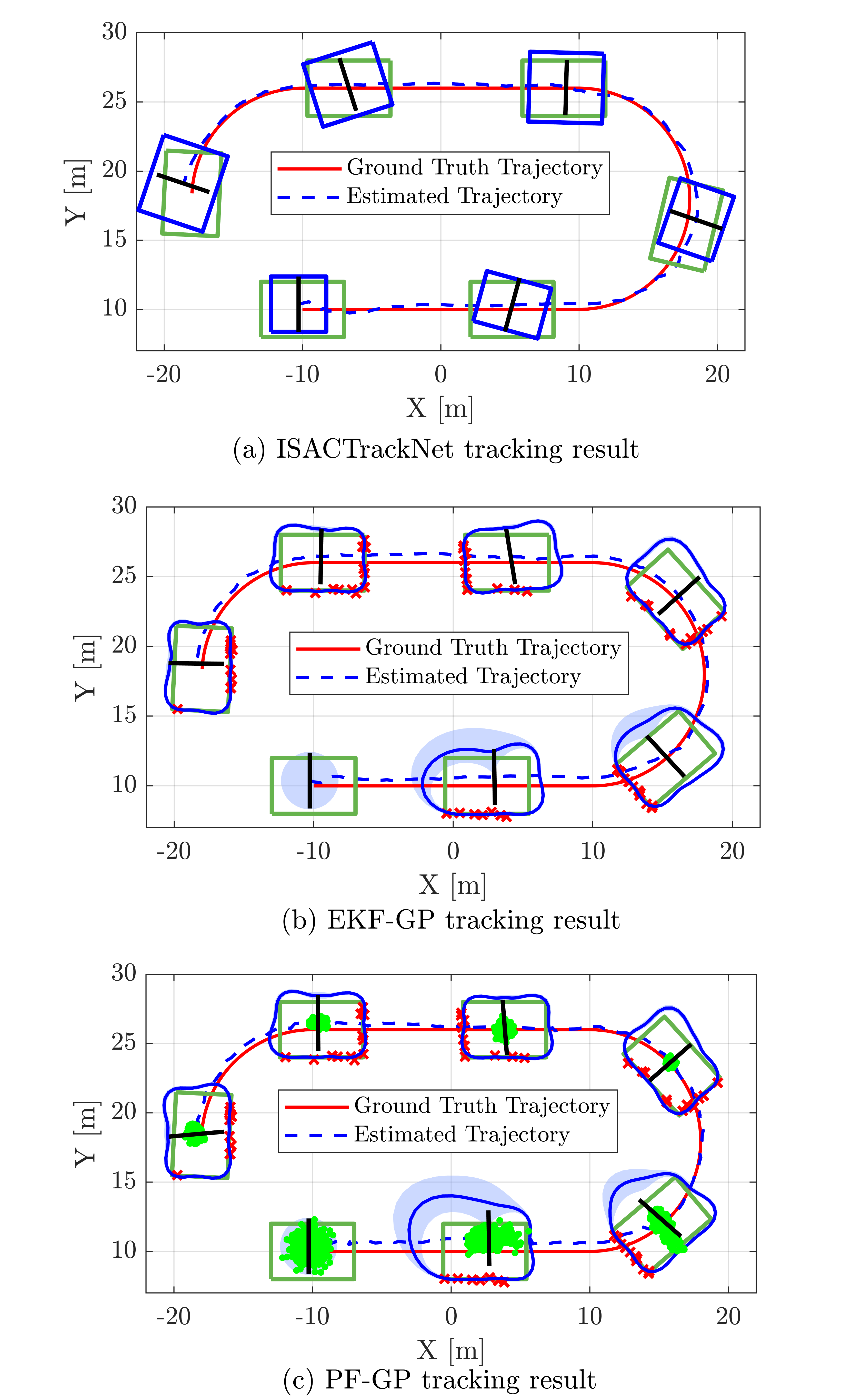}
\caption{Estimated trajectories for different tracking schemes. The ground truth and estimated contours are depicted with green and blue boxes, respectively. The shallow blue area represents the confidence region for EKF-GP and PF-GP schemes. The ET orientation is represented by a perpendicular black line. The red and green markers are respectively the scatterers and PF particles.}
\label{fig_Trajectory}
\end{figure}
\section{Numerical Results}
\subsection{Simulation setup}
1) \textit{System setting}: The BS is equipped with a transmit ULA of $N_t =15$ elements and a receive ULA of $N_r = 15$ elements. The separation between the centers of two ULAs is $2\ \rm{m}$. We set the transmit power as $P_t = 30\ \rm{dBm}$, the signal wavelength to $0.01\ \rm{m}$, and the noise power to $\sigma^2=-80\ \rm{dBm}$. There are $K_{\rm{CL}}=3$ clutters around the BS. The power of residual SI is $p_{\rm{SI}}= -90\ \rm{dBm}$. The rectangular ET has length $L=6\ \rm{m}$ and width $W=4\ \rm{m}$, which is tracked for $N =200$ consecutive time slots with each slot being $T=1\ \rm{s}$.

2) \textit{ISACTrackNet setting}: The DNNs in the Denoising and Encoder modules are composed of multilayer perceptrons. The forgetting factor is set to $\alpha=0.2$. The numbers of training, validation, and test samples are respectively $M=8192$, $2048$, and $1024$, generated based on random U-shaped trajectories, ET size, SI, and clutters. Each long trajectory is divided into multiple small samples of $40$ periods for training convergence \cite{Revach22TSP}. We use a noisy initial ET state in KalmanNet and a square with side length $4\ \rm{m}$ as the initial contour. The training phase is terminated if the losses of the validation set do not decrease within $200$ epochs. The experiments are performed on an Intel Xeon Silver 4214R CPU and a 24\,GB Nvidia GeForce RTX 3090 Ti graphics card with PyTorch of CUDA 11.4.

3) \textit{Benchmark setting}: We select the following two radar-based ET tracking algorithms for comparison, which uses the location measurements of multiple scatterers for ET tracking. These algorithms are generally inapplicable to ISAC systems, since only a limited number of scatterers can be extracted from communication echo sequences.
\begin{itemize}
    \item \textit{Extended Kalman Filter with Gaussian Process (EKF-GP)} \cite{Wahlström15TSP}: It uses GP for ET contour modeling. The target contour is separated into several anchor points which are jointly tracked with other ET states via an EKF.
    \item \textit{Particle Filter with Gaussian Process (PF-GP)} \cite{Emre16FUSION}: This scheme uses the same GP-based contour modeling in \cite{Wahlström15TSP}, but the ET states are instead tracked by a particle filter.
\end{itemize}
\begin{figure}[!t]
\centering
\includegraphics[width=3.38in]{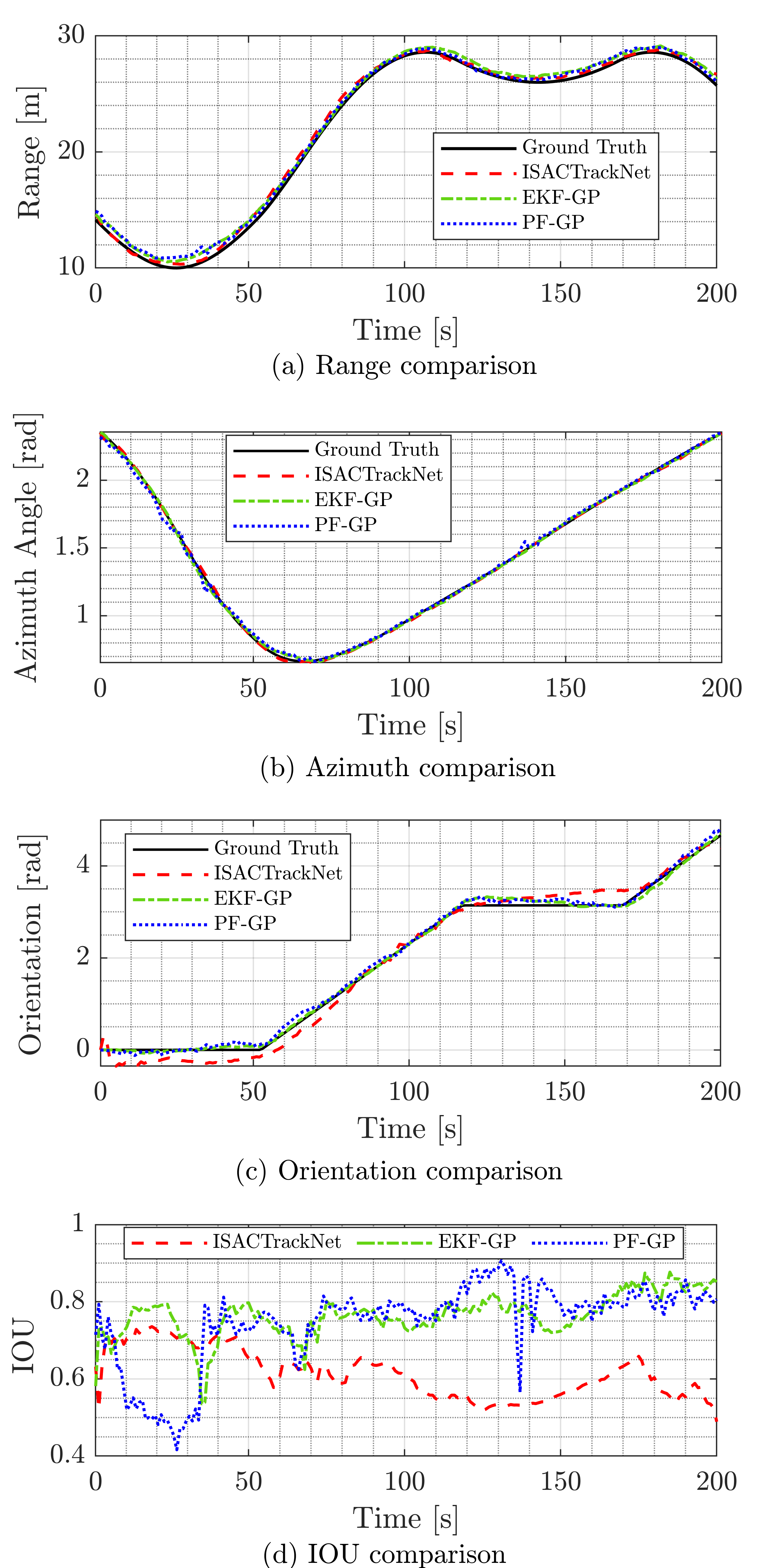}
\caption{Tracking performance comparison for different tracking schemes. The tracked trajectory is the same as in Fig.~\ref{fig_Trajectory}.}
\label{fig_Comparison}
\end{figure}
\subsection{Simulation Results}
The estimated trajectories of different tracking schemes are illustrated in Fig.~\ref{fig_Trajectory}. We observe that despite the presence of maneuvers, such as turns, which deviate from the constant-velocity motion model, all three algorithms effectively capture the overall movement of the ET. Among them, ISACTrackNet demonstrates higher tracking accuracy during linear motion segments compared to EKF-GP and PF-GP. However, its localization performance slightly degrades during rotational motion. Regarding contour estimation, the performance of all aforementioned algorithms gradually converges as the number of observed samples increases. In particular, the GP-based contour modeling struggles to converge when certain edges are occluded and remain unobserved, leading to suboptimal estimation until all edges have been observed. In contrast, ISACTrackNet, which employs a rectangular modeling approach, benefits from its inherent symmetry, allowing it to achieve faster convergence with fewer observations. Nevertheless, the contour alignment of ISACTrackNet remains suboptimal due to accumulated errors in position and orientation estimation.

We further investigate the detailed tracking performance in Fig.~\ref{fig_Comparison}. Here, the intersection-over-union (IOU) shown in Fig.~\ref{fig_Comparison}(d) is used to evaluate the ET extent estimation, defined as the ratio of the areas for the intersection and the union of the estimated/true ET regions \cite{Wahlström15TSP}. From Fig.~\ref{fig_Comparison}(a) and \ref{fig_Comparison}(b), we observe that ISACTrackNet exhibits superior tracking accuracy in terms of target distance and angle. However, it performs slightly worse than EKF-GP and PF-GP (radar-based algorithms) in terms of orientation angle tracking and IOU performance, as shown in Fig.~\ref{fig_Comparison}(c) and \ref{fig_Comparison}(d).

\textcolor{black}{The tracking performance degradation occurs primarily during ET maneuvering phases (e.g., when ET turns at the end of trajectory), due to the underlying constant-velocity model assumption adopted in both KalmanNet within ISACTrackNet and Kalman Filter within the benchmark algorithms. KalmanNet typically exhibits slower adaptation to such model mismatches compared to traditional Bayesian method in the benchmark algorithms. Additionally, the limited number of transceiving antennas restricts the spatial feature extraction capabilities of ISACTrackNet. It is expected that such learning-based ET tracking performance could be significantly improved by employing larger antenna arrays.}

\section{Conclusion}
In this work, we present a deep learning-based ET tracking scheme in an ISAC system, which exploits communication signals to estimate a rectangular ET’s center, orientation, and contour. By integrating a Denoising module for residual SI and clutter suppression, an Encoder module for instantaneous state prediction, and the KalmanNet for iterative state refinement, the proposed ISACTrackNet directly infers the ET state without explicit scatterer-level resolution. Simulation results demonstrate that the proposed method attains rapid convergence and high accuracy across complicated trajectories, achieving near-optimal performance relative to GP-based filters in cluttered scenarios. In future work, more sophisticated ET shape models and multi-target tracking in dynamic environments can be explored to further enhance system scalability and robustness.
\input{Reference.bbl}

\end{document}